%% file: astroph.tex
\input
\documentstyle[12pt,aaspp4,flushrt]{article}
\tighten

\begin{document}

\title {\Large Millimeter-wave 
Aperture Synthesis Imaging of Vega: \\ Evidence for 
a Ring Arc at 95 AU}
\author {D.W. Koerner\altaffilmark{1}, A.I. Sargent\altaffilmark{2}, 
and N.A. Ostroff\altaffilmark{1}}
\altaffiltext{1} 
{University of Pennsylvania, David Rittenhouse Laboratory, 
209 S. 33rd St., Philadelphia, PA 19104-6396}
\altaffiltext{2}
{Division of Physics Mathematics and
Astronomy, California Institute
of Technology, Pasadena, CA 91125}

\bigskip
\bigskip

\begin{abstract}

We present the first millimeter-wave aperture synthesis map
of dust around a main sequence star. A 3$''$ resolution image 
of 1.3 mm continuum emission from Vega reveals a clump of
emission 12$''$ from the star at PA 45$^\circ$,
consistent with the location of maximum 850 $\mu$m
emission in a lower resolution JCMT/SCUBA map.
The flux density is 4.0$\pm$0.9 mJy. Adjacent 1.3 mm
peaks with flux densities 3.4$\pm$1.0 mJy and 2.8$\pm$0.9 mJy
are located 14$''$ and 13$''$ from the star
at PA 67$^\circ$ and 18$^\circ$, respectively. An arc-like bridge
connects the two strongest peaks. 
There is an additional 2.4 $\pm$0.8 mJy peak to the SW 11$''$ 
from the star at PA 215$^\circ$ and a marginal detection,
1.4$\pm$0.5 mJy, at the stellar position, consistent with 
photospheric emission. 
An extrapolation from the 850 $\mu$m flux, assuming
$F_{1.3mm-0.85mm} \propto \lambda^{-2.8}$, agrees well
with the total detected flux
for Vega at 1.3 mm, and implies a dust emissivity index, 
$\beta$, of 0.8 We conclude that we have detected all but a
very small fraction of the dust imaged by SCUBA in our aperture
synthesis map and that these grains are largely confined to
segments of a ring of radius 95 AU.

\end{abstract}

\section {Introduction}

The IRAS detection of excess infrared emission
from Vega ($\alpha$ Lyrae) heralded the discovery
that dust grains surround many nearby stars
(Aumann et al.\ 1984; Gillett 1986). 
High-resolution images at wavelengths ranging from the optical
to millimeter regimes have since revealed that, in many cases, 
the dust around similar ``Vega-excess'' stars
is configured in circumstellar disks
and rings (Smith \& Terrile 1984; see Koerner 2001 for a review).
Some of the more massive disks surround stars with ages of
order 10 Myr (e.g., Smith \& Terrile 1984; Jura 1991; Zuckerman
\& Becklin 1993a) and are readily explainable as the end stages
of protostellar/protoplanetary disks like
those around younger T Tauri stars (e.g., Beckwith \& Sargent 1993). 
The persistence of material in orbit
around older stars like Vega (age $\sim$350 Myr; 
Maeder \& Meynet 1988) is more 
problematic,  since radiation pressure and Poynting-Robertson
drag should have dispersed any grains on much shorter 
timescales. Traditionally, this  conundrum has been 
resolved by postulating an unseen population of asteroidal or
cometary bodies that collisionally regenerate a ``debris disk''
(e.g., Backman \& Paresce 1993), although
particles may also be dynamically trapped in mean motion
resonances
with larger orbiting bodies (Ozernoy et al.\ 2000).

Coronagraphic imaging of $\beta$ Pic quickly provided a 
clear picture of its circumstellar disk (Smith \& Terrile 1984). 
However, 
it has proved difficult to establish unambiguously
the distribution of dust around Vega.  The first IRAS observations 
resolved the 60 $\mu$m emission in a 25$''$ beam and 
indicated a 60-100 $\mu$m color temperature of 85 K (Aumann et al.\ 1984). 
Gillett (1986)  estimated a size of $29''\times25''$  (FWHM)
at PA 85$^\circ$ and postulated the presence of
a dust reservoir radially confined between 36 and 170 AU for large grains
($> 100 \mu$m),
or between 14 and 400 AU for small grains ($< 10 \mu$m).
He also noted that grains smaller than 
1.1 mm should have been depleted by Poynting-Robertson 
drag. A re-examination of the original resolved 60 $\mu$m
emission led to a source size of 27$''$ $\times$ 27$''$ (FWHM), 
implying a radius of 106 AU and grains ranging in size from 0.1 
to 10 $\mu$m (van der Bliek et al.\
1994). A pole-on configuration was inferred from the circular
symmetry and is consistent with the stellar orientation
implied by optical spectroscopy (Gulliver et al.\ 1994). 
Kuiper Airborne Observatory (KAO) measurements by
Harvey et al.\ (1984), in 30$''$ and 43$''$ beams at 
$\lambda$ = 47 and 95 $\mu$m, yielded a lower color temperature,
78 K, than IRAS, and suggested that
different amounts of material are enclosed by the different beams. 
Combining these results with measurements in KAO 
apertures placed 1$'$ away from the star, 
these authors proposed that the extent of the dust around Vega
was as large as 46$''$(FWHM). Recent ISOCAM 
observations at 60 and 90 $\mu$m support this suggestion,
yielding Gaussian widths (60\% peak value)
of 22$''\pm2''$ and 36$''\pm3''$, respectively
(Heinrichsen et al.\ 1998). Deconvolution of the 
60 $\mu$m ISOCAM image was found to be consistent with
an extended source surrounded by a ring
of 42$''$ diameter. ISOPHOT measurements from
25 to 200 $\mu$m imply a color temperature of 73 K,
a grain emissivity index, $\beta = 1.1$, where emissivity 
$Q \propto \lambda^{-\beta}$,
and the presence of grains of size 
110 $\mu$m (Heinrichsen et al.\ 1998). 

Initial observations of Vega at sub-millimeter wavelengths were 
carried out using beams of diverse sizes which complicated interpretation 
(Chini et al.\ 1990; 1991; see discussion in  Zuckerman \& Becklin 1993b). 
A seven-point map at $\lambda = 800 \mu$m,
extending to 16$''$ from the star, 
revealed excess emission in an 18$''$ diameter field
8$''$ E and 14$''$ N of the stellar position (Zuckerman \& Becklin 1993b).
This first indication of asymmetric morphology also
demonstrated that the source extended beyond many of the beams used in
early submillimeter-wave measurements.

The asymmetric distribution of dust was dramatically confirmed by
Holland et al.\ (1998) (hereafter HGZWM)
in their 850 $\mu$m image of Vega obtained
at the JCMT using the SCUBA bolometer array. Although 
extended emission appears circular at the lowest intensity
levels in this image, a more linear core oriented at PA $\sim$45$^\circ$ 
encompasses a bright peak at 9$''$  (70 AU) from the 
estimated position of the star.  Combining the SCUBA results with Vega 
fluxes from 10$\mu$m to 1 mm, Dent et al.\ (2000) (hereafter, DWHG)
find a dust temperature of 80 K, effective grain sizes of 70 $\mu$m, and
$\beta$ = 0.8. Their modelling suggests that the dust is
confined to a ring centered on the star with inner radius $>$ 80 AU
and outer radius $<$ 120 AU (i.e., between
$10''$ and $15''$ from the star). The brightness asymmetry seen in the SCUBA
maps demands density enhancements of at least a factor of 2 in the NE 
sector of the putative ring. With the higher resolution attainable with 
the Owens Valley Radio Observatory (OVRO) millimeter-wave array, we can
image the distribution of dust grains around Vega in greater detail.
Here we present 3$''$ resolution aperture synthesis images of 1.3 mm
emission from Vega that strongly support the pole-on ring hypothesis.

\section{Observations and Results}

Two low-resolution configurations 
(baseline lengths from 15 to 115 m) 
of the six 10.4 m telescopes of the OVRO millimeter array 
were used 
to observe Vega in the 1.3 mm band in four 1 GHz bands 
at 229.0, 230.5, 233.5, and 235.0 GHz. Observations were made 
on 15 occasions between fall 1999 and spring 2001.
These observing cycles each lasted from 4 to 10 hours and included 
25 minute integrations on Vega interspersed with 10 minute
integrations on 3C345 for phase and amplitude visibility
calibration. The flux density for 3C345 was based on
measurements of Neptune and Uranus and  
varied between 4.0 and 4.6 Jy during the course
of the project. The phase center of the observations
was the optical position of Vega, {\sl $\alpha(1950)\  8^h 35^m 14.655^s, \
\delta(1950) +38^\circ 44' 09.68''$ }, modified by proper motions  
$\Delta\alpha $ = 201 mas/yr and $\Delta\delta$ = 285 mas/yr.
These proper motions resulted in  
a total positional shift of less than 0.5$''$ throughout the
observing period. Visibilities
were calibrated using Caltech's MMA software and were combined
with NRAO's AIPS package. The AIPS task, IMAGR, was used
to make a map which combined natural weighting with
a Gaussian taper. A taper with half-power radius of 80 k$\lambda$ 
yielded the best signal to noise ratio.
The resulting synthesized beam size is $3.3'' \times 2.9''$ (FWHM)
at PA 119$^\circ$ and the rms noise amplitude is 0.5 mJy.

Continuum emission is detected at several locations 
in the 1.3 mm image of Vega displayed in Fig.\ 1a. For each,
the separation from the star and flux density is  
listed in columns 1-3 of Table 1.
The brightest peak lies $12.2''\pm0.6''$ NE of the stellar
position (95 AU at the 7.8 pc distance of Vega) and is flanked by 
two fainter detections. The brighter of these is 
linked to the main peak by an arc of low-intensity emission.
The location of the 1.3 mm maximum is consistent with
that of the peak of 850 $\mu$m emission in a 14$''$ beam
found by HGZWM, $9''\pm2''$ NE of the star. 
An additional weaker source is seen 
SW of the star at a 
distance of $11''\pm1''$ (86 AU). Its location and 
brightness relative to the NE intensity enhancements
are also consistent with the morphology of the SCUBA map.
Our marginal detection of radiation at the stellar position,
1.4$\pm0.5$ mJy, is compatible with the expected photopheric
value, 2.5$\pm1.0$ mJy, extrapolating from model fluxes at shorter
wavelengths (Cohen et al.\ 1992).

Since the emission peaks in our map are not far from the half-power
radius of the primary OVRO array beam, $\theta_B/2 = 16''$, 
their brightness is under-represented in Fig.\ 1a. 
Correction for this diminution was carried
 out with AIPS task PBCOR and produced the map
displayed in Fig.\ 1b. The corresponding flux densities are
listed in column 4 of Table 1. For comparison with our image,
the 850 $\mu$m fluxes (HGZWM) must be
extrapolated with an appropriate relation, $F_\lambda \propto 
\lambda^{-\alpha}$. These extrapolated values are tabulated in Table 1,
columns 5 and 6, for $\alpha$ = 2 and 3 respectively. Corresponding
SCUBA fluxes are listed in Column 7. For
$\alpha$ = 2, the fluxes listed in column 4 
account for only 75\% of the 850$\mu$m flux; with $\alpha$ = 3,
the OVRO and SCUBA fluxes agree within the uncertainties.

\section{Discussion}

The 3$''$ resolution OVRO maps in Fig.\ 1 
strongly support the hypothesis that the dust around
Vega is distributed in a clumpy ring, 
as DWHG suggested. 
Emission appears at distances of 85 to 110 AU from the
star. Its brightness and distribution 
is consistent with the size and asymmetry of the 
emission in the 14$''$-resolution SCUBA map.
The brightest 1.3 mm clump, at 95 AU, is connected to 
a neighboring peak by a low-intensity bridge which forms an arc-like 
structure centered on the star. Dust is also detected 
on the opposite side of the star at a nearly equidistant position.  
All the flux from the SCUBA map is recovered in the 1.3 mm detections,
if a plausible emissivity law is adopted. This suggests
that there is little broad extended emission. However,
our observations do not rule out the presence
of material beyond the edge of the OVRO primary beam, 
16$''$ (125 AU), as suggested by 95 and 90 $\mu$m
measurements (Harvey et al. 1984; Heinrichsen et al.\ 1998). 

The total flux in the 850 $\mu$m
map of HGZWM, 45.7 mJy, implies a 1.3 mm flux density
equal to that of sources detected in our map, 14 mJy, 
if the wavelength dependence of the flux density 
$F_{0.85-1.3mm} \propto \lambda^{-\alpha}$ has
$\alpha = 2.8$. HGZWM report a value of $\alpha$ = 2.7 
for $F_\lambda$ between 850 $\mu$m and 1350 $\mu$m in
a single aperture centered on the peak of 850 $\mu$m emission.
For optically thin dust radiating in the
Rayleigh-Jeans regime, our best fit value for $\alpha$
implies a grain emissivity, $Q \propto \lambda^{-\beta},$ 
with $\beta = 0.8$.  The same value was derived by DWHG in modeling
of fluxes from 10 to 1000 $\mu$m and, together with the radial
location of the dust, implies grain sizes between
$30$ and $200 \mu$m. 

Taken together, our results are consistent with Vega being surrounded
by a ring of highly variable density. From model fitting, both to
the image of HGZWM and to long-wavelength fluxes, DWHG predicted a
circum-Vega ring with density enhancements of a factor of two in the NE
segment. Emission at levels less than half that of even the strongest
peak in Fig.\ 1a would be below our 3$\sigma$ (1.5 mJy) detection
threshold. As a consequence, a continuous dust ring
remains unseen, but the arc-like structure of 
the patchy maxima provide compelling evidence for its
presence. The detection of a much weaker,
but symmetrically placed, dust continuum peak to the SW
adds to this evidence. Morever, in Fig.\ 1b, after correction
for the effects of the primary beam, several clumps are
detectable at the 2$\sigma$ confidence level. All lie along
a possible orbital trajectory for ring material.

There is some evidence for clumping of dust in rings
in other extra-solar systems. The narrow ring around HR 4796
(Koerner et al.\ 1998;
Schneider et al.\ 1999) shows strong evidence of asymmetric brightening
at one ansa (Telesco et al.\ 2000). This feature has been explained
as the result of eccentricity forcing, either by a stellar companion
or by an unseen planet (Wyatt et al.\ 2000). An 850 $\mu$m SCUBA image of 
$\epsilon$ Eridanae shows a still greater degree of azimuthal
asymmetry (Greaves et al.\ 1999). The ring discontinuity 
presented here is even more pronounced. Images in Fig.\ 1 present 
a picture that resembles that discovered by stellar occultations of
the planet Neptune; partial ``ring arcs'' were inferred from the 
fact that occultations were observed only along preferred trajectories  
(Hubbard et al.\ 1986). Subsequent Voyager images revealed
that the arcs were actually connected by tenuous ribbons of dust
(Smith et al.\ 1989).
Theoretical analyses concluded that the 
material is azimuthally confined by resonant interactions
with Neptune's satellite Galatea (Goldreich et al.\ 1986;
Porco 1991; Hanninen \& Porco 1997). The observations presented here
suggest that there may be useful parallels to be drawn between
the ring-arcs associated with Neptune and the distribution of
dust around Vega. Future sub-mm and mm-wave
observations, especially with SAO's Sub-Millimeter Array (SMA)
and with the Combined Array for Research in Millimeter Astronomy 
(CARMA), should make it possible to determine Vega's dust structure
in even greater detail and infer details of any hidden planetary 
bodies responsible for the morphology of a ring arc.

\section{Acknowledgements}

We are grateful to all OVRO postdoctoral scholars and visiting
astronomers who helped in acquiring the data, but especially 
to Kartik Sheth. 
OVRO millimeter array is supported by NSF through grant AST 998154.
OVRO research on other solar systems is also supported by
Norris Planetary Origins Project and NASA Origins Grant NAG5-9530.

\bigskip

\begin{figure}
\plotone{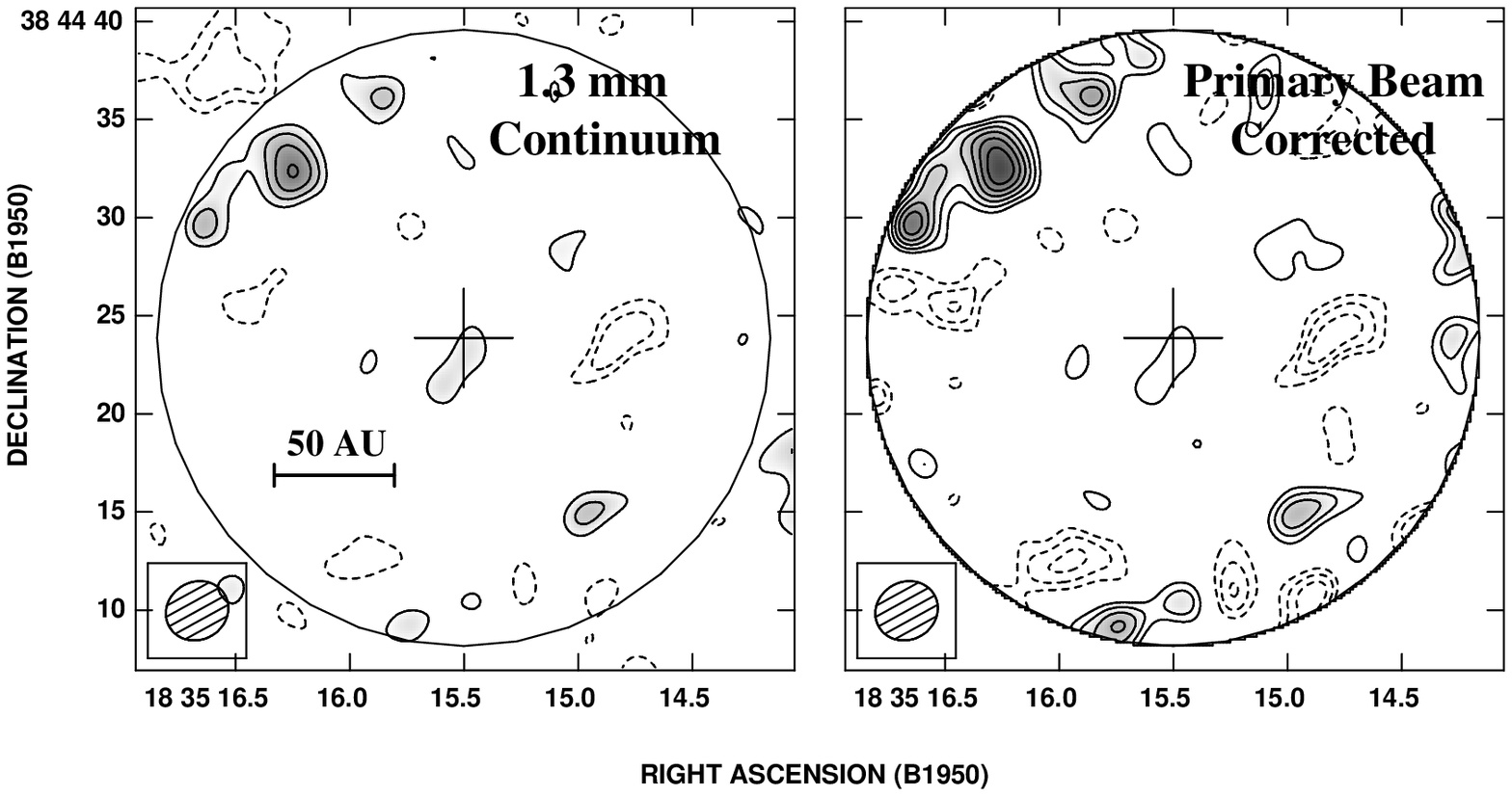}

\figcaption{{\bf Left:} Aperture synthesis image of 1.3 mm emission
from Vega ($\alpha$ Lyrae). Contours are at 1$\sigma$ intervals of 0.5 mJy,
starting at the 2$\sigma$ level. The position of Vega is marked
with a cross, the half-power circumference of the primary beam of
the 10.4 m telescopes is 
shown as a circle. The synthesized beam, $3.3'' \times 2.9''$ FWHM
at PA 119$^\circ$, is represented by a hatched circle. 
{\bf Right:} Same as at left, except that 
compensation has been made for radial fall-off in intensity
due to the primary beam pattern.}

\end{figure}

\bigskip

\input{fluxtab.tex}

\vfill

\end{document}

%% file: fluxtab.tex
\singlespace
{\small
\begin{deluxetable}{ccccccc}
\tablenum{1}
\tablewidth{6.0truein}
\tablecaption{Flux Densities in Map}
\tablehead{
\colhead{(1)} & \colhead{(2)} & \colhead{(3)} &
\colhead{(4)} & \colhead{(5)} & \colhead{(6)} & \colhead{(7)} \\
\colhead{Separation} & \colhead{PA} & \colhead{$F_{1.3 mm}$} & \colhead{$F_{1.3 mm}$} &
\colhead{$F_{0.85 mm}$} & \colhead{$F_{0.85 mm}$} & 
\colhead{$F_{0.85 mm}$} \\
\colhead{(arc sec) } & \colhead{(degrees)} 
&\colhead{(mJy)} & \colhead{corrected} & 
\colhead{$\alpha = 2$} & \colhead{ $\alpha = 3$} &  
\colhead{(SCUBA)} 
}

\startdata
12.2$\pm$0.6 & 45 & 2.5$\pm$0.6 & 4.0$\pm$0.9 & 9.4 & 14.3 & 17.3$\pm$3    \\
14.5$\pm$1.0 & 67 & 1.8$\pm$0.5 & 3.4$\pm$1.0 & 8.0 & 12.2 &  --   \\
13.0$\pm$1.0 & 18 & 1.7$\pm$0.5 & 2.8$\pm$0.9  & 6.6 & 10.0 & --    \\
11.0$\pm$1.0 & 215 & 1.7$\pm$0.5  & 2.4$\pm$0.8 & 5.6 & 8.6 &  $\sim$15 \\
0 & --  & 1.4$\pm$0.5 & 1.4$\pm$0.5 & 5$\pm$1$^a$ & 5$\pm$1   & --  \\

\hline

Total & -- & 9.1$\pm$1.2 & 14$\pm$1.9 & 32.9 & 48.4    & 45.7$\pm$5.4$^b$ \\
\enddata


\tablenotetext{a}{Represents estimated photospheric flux from Vega at
$\lambda$ = 850$\mu$m.}
\tablenotetext{b}{Integrated emission from map of Holland et al.\ 1998.}

\end{deluxetable}
}
\doublespace

%% file: astroph.bbl
\begin{references}

\reference{aum}
Aumann, H.H., Gillett, F.C., Beichman, C.A., de Jong, T., Houck, J.R.,
Low, F.J., Neugebauer, G., Walker, R.G., \& Wesselius, P.R.\ 1984,
\apj, 278, L23.

\reference{bp} Backman, D.E., \& Paresce, F.\ 1993, in Protostars \& 
Planets III, ed. E.H. Levy \& J.I. Lunine, 
(Tucson: University of Arizona Press), p.\ 1253

\reference{bsp3} Beckwith, S.V.W., \& 
Sargent, A. I.\ 1993, in {\it Protostars and
Planets III}, eds.\ E.H. Levy \& J.I. Lunine
(Tucson:University of Arizona Press), p.\ 521 

\reference{coh}
Cohen, M., Walker, R.G., Barlow, M.J., \& Deacon, J.R.\ 1992, 
AJ, 104, 1650

Chini, R., Kr\"ugel, E., \& Kreysa, E.\ 1990, A\&A, 227, L5
\reference{chini1}
Chini, R., Kr\"ugel, E., \& Kreysa, E.\ 1990, A\&A, 227, L5

\reference{chini2}
Chini, R., Kr\"ugel, E., Shustov, B., Tutukov, A., 
\& Kreysa, E.\ 1991, A\&A, 252, 220

\reference{dent}
Dent, W.R.F., Walker, H.J., Holland, W.S., \& Greaves, J.S.\ 
2000, MNRAS, 314, 702 (DWHG)

\reference{gill}
Gillett, F.C.\ 1986, in {\it Light on Dark Matter}, ed.\ F.P. Israel, 
(Dordrecht: Reidel), p.\ 61

\reference{peter}
Goldreich, P., Tremaine, S., \& Borderies, N.\ 1986, AJ, 92, 490

\reference{als}
Greaves, J.S., Holland, W.S., et al.\ 1998, ApJ, 506, 133

\reference{gull}
Gulliver, A.F., Hill, G., \& Adelman, S.J.\ 1994, \apj, 429, L81 

\reference{hp}
Hanninen, J., \& Porco, C.\ 1997, Icarus, 126, 1

\reference{harv}
Harvey, P.M., Wilking, B.A., \& Joy, M.\ 1984, Nature, 307, 441

\reference{hwk}
Heinrichsen, I., Walker, H.J., \& Klaas, U.\ 1998, MNRAS, 293, L78

\reference{hol} Holland, W.S., Greaves, J.S., Zuckerman, B., Webb, R.A.,
McCarthy, C., Couldon, I.M., Walther, D.M., Dent, W.R.F.,
Gear, W.K., \& Robson, I.\ 1998, \nat, 382, 788 (HGZWM)

\reference{hub}
Hubbard, W.B., Brahic, A., Sicardy, B., Elicer, L.-R., 
Roques, F., \& Vilas, F.\ 1986, Nature, 319, 636

\reference{Jura1} Jura, M.\ 1991, \apj, 383, L79

\reference{teton}
Koerner, D.W.\ 2001, in {\it The 4th Teton Summer School Conf.: Galactic Structure, Stars, and the Interstellar Medium,}  ed. C. W. Woodward, M. Bicay, \& 
J. M. Schull, (San Francisco: ASP), p.\ 563
 
\reference{als}
Koerner, D.W., Ressler, M.E., Werner, M.W., \& Backman, D.E.\ 1998, 
ApJ, 503, L83

\reference{MM}
Maeder, A., \& Meynet, G.\ 1988, A\&AS, 76, 411

\reference{oz}
Ozernoy, L.M., Gorkavyi, N.N., Mather, J.C., Taidakove, \&
T.A.\ 2000, \apj, 537, L147

\reference{cp1}
Porco, C.C.\ 1991, Science, 253, 995


\reference{S99} Schneider, G., Smith, B.A., Becklin, E.E., Koerner, D.W.,
Meier, R., Hines, D.C., Lowrance, P.J., Terrile, R.J., Thompson, R.I.,
\& Rieke, M.\ 1999, \apj, 513, L127


\reference{sft} 
Smith, B.A.,  \& Terrile R.J.\ 1984, Science, 226, 4681


\reference{smi}
Smith, B.A., Soderblom, L.A., Banfield, D., Barnet, C., Beebe, R.F.,
Bazilevskii, A.T., Bollinger, K., Boyce, J.M., Briggs, G.A.,
\& Brahic, A.\ 1989, Science, 246, 1422

\reference{tel}
Telesco, C.M., Fisher, R.S., Piña, R.K., Knacke, R.F., Dermott, S.F., 
Wyatt, M.C., Grogan, K., Holmes, E.K., Ghez, A.M., Prato, L., 
Hartmann, L.W., \& Jayawardhana, R.\ 2000, \apj, 530, 329


\reference{van}
van der Bliek, N.A., Prusti, T., \& Waters, L.B.F.M.\ 1994, 
A\&A, 285, 229


\reference{wy}
Wyatt, M.C., Dermott, S.F., Telesco, C.M., Fisher, R.S., Grogan, K.,
Holmes, E.K., \& Pi\~na, R.K.\ 2000, \apj, 527, 918

\reference{zba}
Zuckerman, B., \&  Becklin, E.E.\ 1993a, \apj, 406, L25

\reference{zbb}
Zuckerman, B., \&  Becklin, E.E.\ 1993b, \apj, 414, 793


\end{references}
